\documentclass[aps,prx,reprint,superscriptaddress]{revtex4-2}
\pdfoutput=1
\usepackage[utf8]{inputenc}
\usepackage[english]{babel}
\usepackage[T1]{fontenc}
\usepackage{amsmath}
\usepackage{hyperref}
\usepackage{tikz}

\begin{document}

\title{A non-Markovian approach to spin-phonon coherence and the breakdown of the Markovian approximation
}

\author{Mohamed Belhassen}
\affiliation{Department of Physics, Humboldt-Universit\"{a}t zu Berlin, Newtonstr. 15, 12489 Berlin, Germany}

\author{Tim Schr\"{o}der}
\email[Corresponding author: ]{tim.schroeder@physik.hu-berlin.de}
\affiliation{Department of Physics, Humboldt-Universit\"{a}t zu Berlin, Newtonstr. 15, 12489 Berlin, Germany}
\affiliation{Ferdinand-Braun-Institut, Gustav-Kirchhoff-Straße 4, 12489 Berlin, Germany}

\author{Gregor Pieplow}
\affiliation{Department of Physics, Humboldt-Universit\"{a}t zu Berlin, Newtonstr. 15, 12489 Berlin, Germany}
\begin{abstract}
Qubit coherence is an essential figure of merit for quantum information processing applications such as quantum computing, or quantum repeaters. Understanding the coherence properties of the underlying physical qubits that facilitate such applications is therefore are often limited by coupling to lattice phonons, which in turn constrains operation temperature. Here we study phonon induced electronic spin decoherence in group-IV vacancy centers in diamond. We begin by modeling the spin–phonon interaction and then employ the widely used Born-Markov approximation, highlighting its inconsistencies in this setting and its deviations from experimental observations. To close the gap between theoretical predictions and experimental results, we relax certain approximations, investigate their contributions to the predicted coherence times, and identify the dominant sources of discrepancy. We further demonstrate that experimentally measured electronic spin coherence dynamics are consistently captured within a non-Markovian framework, and we show how the magnetic field orientation influences the qubit coherence time.
\end{abstract}
\maketitle

\section{Introduction}

Group-IV vacancies (G4V) in diamond have enabled demonstrations of metropolitan-scale quantum networks \cite{pezzagna2021,ruf2021}, blind quantum computing \cite{wei2025}, and entanglement-enhanced sensing \cite{knaut2024entanglement}. They are actively explored as qubit platforms for quantum computing \cite{pezzagna2021} and may offer a route to scalability and error correction, for example through modular architectures \cite{singh_modular_2025, abobeih2022a}. However, decoherence remains a major challenge, firmly placing current technology in the noisy intermediate-scale quantum regime. The field is still developing an optimal combination of physical qubits with sufficiently long coherence times and quantum error correction to achieve fully scalable, fault-tolerant quantum computation.

Obtaining a precise understanding of the fundamental limits of qubit coherence at finite temperature remains an open challenge. This question is not only of fundamental interest, but also directly impacts the viability of a given defect and a host-material platform. Even temperature differences on the scale of a kelvin can determine whether a spin qubit remains sufficiently coherent in a cryogenic setup based on liquid helium \cite{zotero-item-5989,harris2023} or instead requires operation close to millikelvin temperatures in a much more complex dilution refrigerator \cite{becker2018}.
\begin{figure}[ht!]
        \centering
        \includegraphics[width=1\columnwidth]{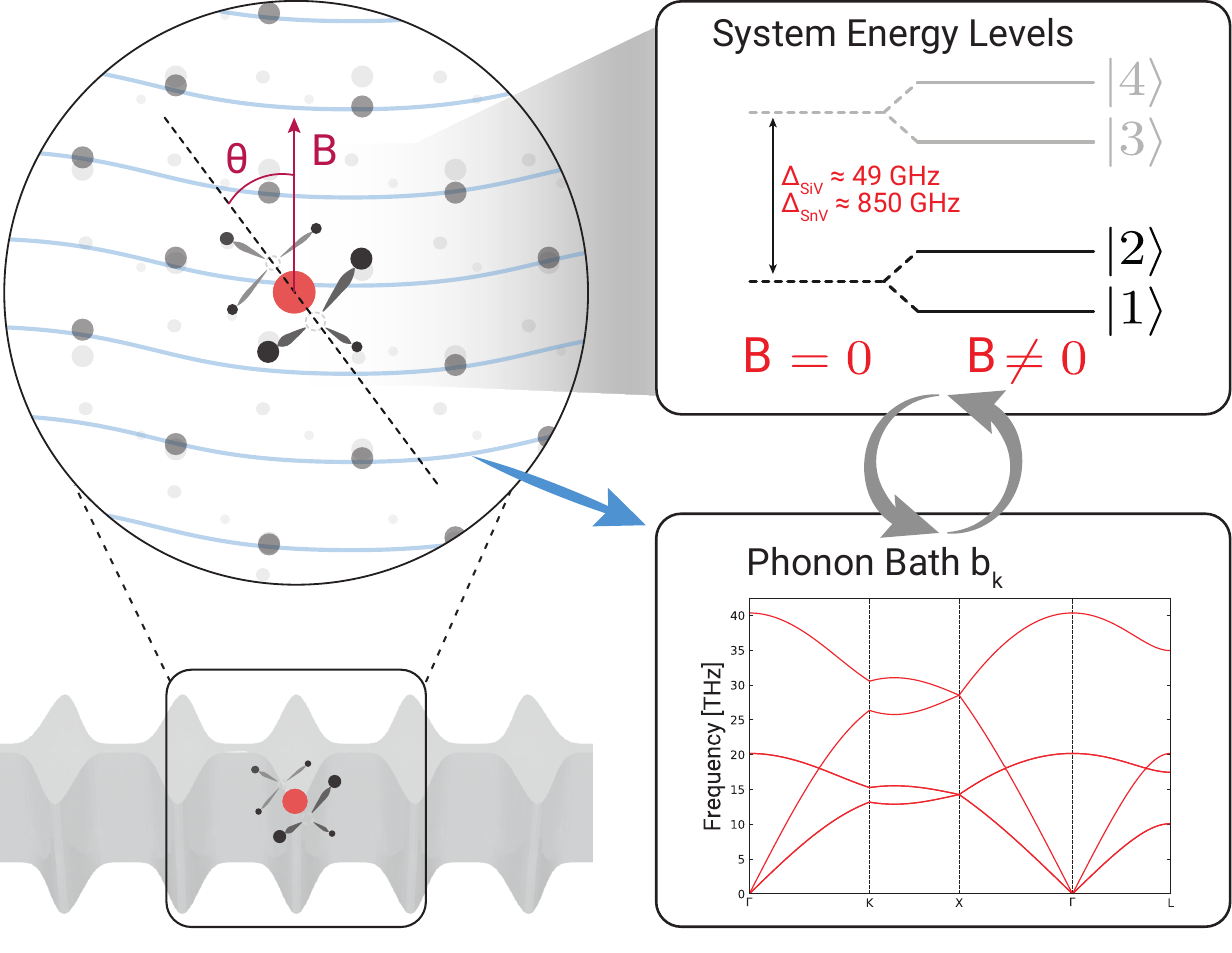}
        \caption{Group IV vacancy in a sawfish resonator \cite{bopp2024}. $b_k$'s are the phonon bath modes, with dispersion relation in bulk diamond plotted in the lower right inset, interacting with the color center energy levels described in the Hamiltonian Eq.~\eqref{eq: unperturbed system}. The magnetic field $B$ is applied with an angle $\theta$ with respect to the color center symmetry axis and splits the orbital spin degeneracy.}
        \label{fig:SnV-Th-Exp T_2}
    \end{figure}
Here we advance the microscopic understanding of electronic spin coherence in G4V. We argue that a quantitative description of their coherence requires a non-Markovian treatment of the solid-state environment together with accurate knowledge of the phonon spectral density. Coherence times predicted by the commonly used Markovian Lindblad master equation can differ substantially from experimental results, with reported discrepancies of up to $20\%$ \cite{harris2023}. In particular, we find that the Markovian approximation underestimates experimentally observed electronic spin coherence times. This discrepancy motivates a non-Markovian treatment and a reexamination of the approximations underlying standard derivations. Furthermore, we account for the vector nature of the magnetic field used to lift the spin-level degeneracy and investigate how the spin-phonon interaction depends on its orientation relative to the defect symmetry axis, which has been shown to strongly affect color-center controllability \cite{pieplow2024}. The vector dependence of the spin-phonon coupling may provide direct information about the phonon spectral density. Magnetic-field-angle-dependent coherence measurements could therefore offer a simple test for ruling out specific modeling assumptions. 

After discussing the intricacies of the spin-phonon model and its Markovian treatment, we go beyond the Markovian approximation by describing phonon-induced non-Markovian dynamics using the Nakajima-Zwanzig and time-convolutionless projection-operator formalisms \cite{breuerTheoryOpenQuantum2009}. We provide a quantitative framework for predicting color-center coherence times once the phonon spectral density is known. By explicitly accounting for system-bath memory effects, we show that the predicted coherence times have higher values relative to Markovian treatments and more closely match experimental observations. This establishes a basis for modeling spin coherence in nanophononic and photonic environments, which is essential for engineered quantum devices.

\section{Spin-Phonon model}

We briefly summarize the spin-phonon interaction model underlying our Markovian and non-Markovian analyses. This summary includes a discussion of the approximations involved and the characteristic energy scales in which particular modeling approaches are valid. This overview is important for placing the regime considered in this work in the context of the current state of the art in modeling the phonon bath of color centers. 

We adopt the spin Hamiltonian of
Ref.~\cite{heppElectronicStructureSilicon2014},
\begin{equation}
    H = H_0 + H_{SO} + H_{JT} + H_{S} + H_B~,
\label{eq: unperturbed system}
\end{equation}
where $H_0$ determines the energies of the unperturbed system,
$H_{\mathrm{SO}}$ denotes the spin-orbit interaction,
$H_{\mathrm{JT}}$ the Jahn-Teller interaction, and $H_{\mathrm{S}}$ the
interaction with strain induced by lattice deformations arising from internal
or external forces. $H_B$ denotes the Zeeman interaction, which lifts the
spin degeneracy in the presence of a static magnetic field. We employ the
same notation and parameter set as in our previous work~\cite{pieplow2024}.

To analyze the effect of the phonon bath, we briefly review the
spin-phonon interaction Hamiltonian obtained from group
theory~\cite{heppElectronicStructureSilicon2014}. 

Symmetry arguments determine the form of the Jahn-Teller and strain
interaction Hamiltonians for group-IV color centers. Phonons may be
described either by projecting lattice vibration modes onto the defect
eigenmodes, yielding a Jahn-Teller-effect-based description, or by
interpreting them as time-dependent compression, tension, and shear
fields, i.e., as a dynamical strain acting on the electronic levels. In
this section, we summarize how these two descriptions give rise to two
distinct spin-phonon coupling models. To the best of our knowledge, the
qualitative differences between these approaches have not been discussed in the diamond color center literature.

{\bf Jahn-Teller-effect-based description}: According to Ref.~\cite{heppElectronicStructureSilicon2014}, the
vibrational motion of the atoms constituting group-IV color centers can
be decomposed into the following modes:
\begin{equation}
    \Gamma_{\text{vib}} = 2A_{1g} + 2E_g + A_{1u} + 2A_{2u} + 3E_u
\end{equation}
where $A$ and $E$ are the 1-dimensional and 2-dimensional irreducible representation respectively. The indices $g$ and $u$ describe parity under inversion symmetry, $g$ for parity $1$ and $u$ for parity $-1$.
 
Only $2A_{1g} + 2E_g$ modes interact with the spin system, and the Hamiltonian of a $A_{1g} + E_g$ type interaction has the form
\begin{align}
\label{eq:Jahn-Teller}
H_{JT}
&= \frac{1}{2}K \left(Q_x^2 + Q_y^2\right) + F
\begin{pmatrix}
Q_x & Q_y \\
Q_y & -Q_x
\end{pmatrix} \nonumber\\
&\quad + G
\begin{pmatrix}
Q_x^2 - Q_y^2 & 2Q_x Q_y \\
2Q_x Q_y & -(Q_x^2 - Q_y^2)
\end{pmatrix}
\end{align}
where $Q_x$ and $Q_y$ are the normal coordinates of the atoms and $K$, $F$ and $G$ are the coupling constants. The first term corresponds to the $A_{1g}$ interaction and can be neglected as we restrict our consideration to the ground-states, which are shifted by an equal amount. The second and third term represent the first- and second-order coupling of the atomic vibrations to the spin degree of freedom.

Note that Eq.~\eqref{eq:Jahn-Teller} gives the Hamiltonian the form of an $A_{1g} + E_g$ type interaction. In the standard treatment \cite{heppElectronicStructureSilicon2014}, group-IV vacancy centers have two copies of these modes. The coupling constants associated with these two copies are indistinguishable in experiment. In addition, strain contributions cannot be distinguished experimentally from the Jahn-Teller interaction, so the relevant coupling constants are not readily available for simulations of the Markovian and non-Markovian dynamics. While these coupling constants could in principle be estimated numerically, doing so even for the two copies of the modes is already computationally very expensive \cite{londero2018} and exceeds our available resources.
We therefore abandon this approach in favor of an alternative one. At comparatively low phonon frequencies, their wavelength is much larger than the size of the color center, so the interaction can be modeled effectively in terms of strain, which we refer to as the low frequency regime.

\textbf{Strain-based model:} Strain describes the deformation of the lattice due to compressive, tensile, or shear forces applied to the diamond and is described by the Hamiltonian \cite{meesala2018,trusheim_transform-limited_2020}
\begin{equation}
\label{eq:Strain}
    H_S = \epsilon_{E_{gx}} \sigma_z \otimes I + \epsilon_{E_{gx}} \sigma_x \otimes I
\end{equation}
where 
\begin{equation}
\begin{aligned}
\epsilon_{E_{gx}} &= d (\epsilon_{xx} - \epsilon_{yy}) + f \epsilon_{zx} \\  
\epsilon_{E_{gy}} &= -2d\epsilon_{xy} + f\epsilon_{yz}  
\end{aligned}
\end{equation}
$\epsilon_{xx}$, $\epsilon_{yy}$, $\epsilon_{zx}$, $\epsilon_{xy}$and $\epsilon_{yz}$ are the components of the strain tensor, $d$ and $f$ are the coupling constants \cite{meesala2018,trusheim_transform-limited_2020}. Comparing Eq.~\eqref{eq:Jahn-Teller} and Eq.~\eqref{eq:Strain}, we find that the Jahn-Teller coupling involves both first- and second-order terms in the atomic coordinates, whereas the strain interaction includes only second-order terms. This shows that the dynamic coupling between the spin and lattice deformation differs from the static coupling, which means that a strain-based phonon model is inherently limited to the low-frequency regime.

Following \cite{harris2023}, we assume that the atom displacement in the lattice can be decomposed into plane modes, which allows us to quantize the strain tensor elements resulting in the spin-phonon system Hamiltonian 
\begin{equation}
    H = H_S + H_B + H_{SB}~,
\label{eq:Total Hamiltonian}
\end{equation}
where $H_S$ is the system Hamiltonian. The bath Hamiltonian is 
\begin{equation}
H_B = \sum_{k m} \hbar \omega_{k m}\,b^\dagger_{k m}\,b_{k m}~,
\end{equation}
where $b^\dagger_{k m}$ and $b_{k m}$ are the phonon creation annihilation operators for the wavevector $k$, with energy $\hbar \omega_{km}$, and polarization $m$. The spin-phonon interaction Hamiltonian is given by
\begin{equation}
H_{SB} = \sum_{\alpha}\hat{\sigma}_\alpha\,B_\alpha
\end{equation}
where $\hat{\sigma}_{\alpha}$ are the Pauli matrices $\sigma_{\alpha}$ represented in the basis of the system eigenstates with $\alpha \in \{z, x\}$ and 
\begin{equation}
    B_\alpha =  {\rm i}\sum_{k m} g_{k m \alpha}\left(b_{k m} - b^\dagger_{k m}\right)
\end{equation}
\begin{equation}
g_{k m \alpha}
= \sqrt{
\frac{\hbar}{2 V \rho \omega_{k m}}}
\, D_{\mathrm{\alpha uv}}\, k_u\, q_{k m v}
\end{equation}
\begin{equation}
    D_{z}
    =
    \begin{pmatrix}
    d & 0 & f/2 \\
    0 & -d & 0 \\
    f/2 & 0 & 0
    \end{pmatrix}
\end{equation}
\begin{equation}
    D_{x}
    =
    \begin{pmatrix}
    0 & -d & 0 \\
    -d & 0 & f/2 \\
    0 & f/2 & 0
    \end{pmatrix}
\end{equation}
where $V$ is the diamond volume, $\rho$ is the density of diamond and $q_{k m v}$ is the phonon polarization vector. The problem is fully specified by Eq.~\eqref{eq:Total Hamiltonian}. Our next step is to derive the master equation used to simulate the system dynamics.

\section{Master equation}

In this section, we study an open-system methods using Eq.~\eqref{eq:Total Hamiltonian} for estimating coherence times while retaining the essential system complexity. We compare a hierarchy of approximations and make explicit the physical assumptions that determine their validity and their impact on the predicted coherence time. Starting with non-Markovian projection-operator techniques, we consider the Nakajima-Zwanzig (NZ) and time-convolutionless (TCL) approaches \cite{breuerTheoryOpenQuantum2009}, before applying the Born-Markov approximation to obtain a Lindblad master equation.

The following projector
\begin{equation}
\begin{aligned}
    \mathcal{P}\rho &= \operatorname{tr}_B\{\rho\} \otimes \rho_B
       = \rho_S \otimes \rho_B\\
       \mathcal{Q} &= 1 - \mathcal{P}
\end{aligned}
\end{equation}
can be used to derive the NZ master equation \cite{breuerTheoryOpenQuantum2009}
\begin{equation}
    \frac{\partial}{\partial t} \mathcal{P}\rho(t)
= \int_{0}^{t} \! ds \, \mathcal{K}(t,s)\, \mathcal{P}\rho(s)~,
\end{equation}
where $\mathcal{K}(t,s)$ is a memory kernel, $\rho_S$ and $\rho_B$ are the density matrices of the system and the phonon bath at thermal equilibrium, respectively. Inserting the total Hamiltonian,
Eq.~\eqref{eq:Total Hamiltonian}, into the memory kernel and expanding
$\mathcal{K}(t,s)$ to second order in the system-bath coupling
\cite{breuerTheoryOpenQuantum2009}
\begin{equation}
\begin{aligned}
    \mathcal{K}(t,s) &= \mathcal{P}\mathcal{L}(t)\mathcal{Q}\mathcal{L}(s)\mathcal{P}\\
    \mathcal{L}(t)\rho &= -\textrm{i} [ H_I(t),\rho ]
\end{aligned}
\end{equation}
together with the Born approximation,
yields
\begin{equation}
\label{eq:Master equation}
    \frac{d}{dt}\rho_S(t)
= - \int_{0}^{t} ds \, \mathrm{tr}_B
\bigl[ H_I(t), [ H_I(s), \rho(s) ] \bigr]~,
\end{equation}
where $H_I(t)$ is the spin-phonon Hamiltonian in the interaction picture. Eq.~\eqref{eq:Master equation} is an integro-differential equation (IDE), is numerically challenging to integrate, but takes into account both the bath and the system memory. This equation can also be obtained by recursively integrating the von Neumann master equation \cite{breuerTheoryOpenQuantum2009}.

The TCL projection method converts the integro-differential equation into a time-local master equation. Instead of keeping the memory term, which depends on the earlier state $\rho(s)$, it rewrites this contribution as a perturbative expansion in terms of the current state $\rho(t)$. For the interaction considered here, only even orders in the system-bath coupling are non-vanishing. The resulting expression is symmetric under the change of variables $s \rightarrow t-s$, which simplifies the time integration.

The Born approximation assumes that the spin and phonon bath are initially uncorrelated and that the bath is weakly coupled to the system. This allows us to right the spin-phonon density matrix as
\begin{equation}
    \rho(t) \approx \rho_S(t) \otimes \rho_B~,
\end{equation}
where $\rho_B$ is assumed to remain unperturbed. Under this assumption, all uneven order contributions in the expansion vanish, and the system-bath evolution will depend on the even order terms. Putting this together, we get the TCL master equation up to the second order
\begin{widetext}
\begin{equation}
\label{eq:NonMarkov}
\begin{aligned}
\frac{d}{dt}\rho_S &=
\sum_{\omega,\omega'}\sum_{\alpha,\beta}
e^{ {\rm i}(\omega' - \omega)t}\,
\Gamma_{\alpha\beta}(\omega, t)\,
\Bigl\{
A_\beta(\omega)\rho_S A_\alpha^\dagger(\omega')
-
A_\alpha^\dagger(\omega')A_\beta(\omega)\rho_S
\Bigr\}
+ \mathrm{h.c.} \\
\Gamma_{\alpha\beta}(\omega, t) &=
\int_{0}^{t}\! ds\, e^{ {\rm i}\omega s}
\left\langle B_\alpha(t)^\dagger\,B_\beta(t - s) \right\rangle
\end{aligned}
\end{equation}
\end{widetext}
where $A_{\alpha}(\omega) \equiv \sum_{\varepsilon' - \varepsilon = \omega} \Pi(\varepsilon)\, \hat{\sigma}_\alpha\, \Pi(\varepsilon')$ and $\Pi(\varepsilon)$ is the eigenvector of $H_S$ with the eigenvalue $\varepsilon$. This equation in general does not guaranty the positivity of the density matrix \cite{schaller2008}. Curiously, this can be ameliorated by choosing a rotating frame and applying a rotating wave approximation (RWA) \cite{schaller2008}.

\subsection{Lindblad Form}

The problem is now reduced to the correlation function 
\begin{widetext}
\begin{equation}
\label{eq:Correlation}
    \left\langle B_\alpha(t)^\dagger\,B_\beta(t - s) \right\rangle = \delta_{\alpha\beta}\chi\int_0^\infty dw_k J(w_k)((n(w_k)+1)e^{- {\rm i}w_k s}+n(w_k)e^{ {\rm i}w_k s})
\end{equation}
\end{widetext}

where 
\begin{equation}
\label{eq:Chi}
\begin{aligned}
    \chi& = \sum_m \iint 
\frac{\hbar \left(D_{\alpha uv}\,\hat{k}_u\, q_{k m v}\right)^2}
{16\pi^3 \rho c_{km}^5}
\, d\Omega\\
n(w)& =\frac{1}{e^{\frac{\hbar w}{k_B T}}-1}
\end{aligned}
\end{equation}
where $\hat{k}_u$ is the unit wavevector, $c_{km}$ is the speed of sound in diamond, $J(w)$ is the phonon spectral density, $n(w)$ is the Bose-Einstein distribution, and $d\Omega$ is the integral over the unit sphere. 

For the purposes of this paper we evaluated the $\chi$s for the SiV and the SnV:
\begin{align}
    &\chi_{\rm SiV} = 1.27 \times 10^{-10} GHz^{-2}~,\\
    &\chi_{\rm SnV} = 3.49 \times 10^{-11} GHz^{-2}~.
\end{align}
The Markovian approximation assumes that the bath correlation function \eqref{eq:Correlation} decays faster than any other time scale in the problem, allowing us to continue the time integral \eqref{eq:NonMarkov} to infinity. One can thus simplify the integration using the identiy
\begin{equation}
\label{eq:identity}
    \int_{0}^{\infty} ds\, e^{- {\rm i} (w-w_k) s}
= \pi\,\delta(w-w_k) -  {\rm i}\,P.V.\,\frac{1}{w-w_k}
\end{equation}
where $P.V.$ is Cauchy principal value. The second term is generally divergent and can be absorbed into a renormalization of the measured parameters.
Finally, the Lindblad form of the Markovian master equation can be optained by performing an RWA, which yields
\begin{equation}
\label{eq: Lindblad}
\begin{aligned}
\frac{d}{dt}\rho_S
&=
\sum_{ij} \gamma_{ij}
\left(
\sigma_{ij}\,\rho_S\,\sigma_{ij}^\dagger
-
\frac{1}{2}
\left\{
\sigma_{ij}^\dagger \sigma_{ij},
\rho_S
\right\}
\right) \\
\gamma_{ij}
&=
2\pi \sum_{\alpha}
\left| \hat{\sigma}_{\alpha_{ij}} \right|^2
\chi_{\alpha}
( \epsilon_i - \epsilon_j )^3
n(\epsilon_i - \epsilon_j)
\end{aligned}
\end{equation}
where $\sigma_{ij} = |i\rangle\langle j|$ and we used the spectral density $J(w) = w^d$ derived from Debye model with $d = 3$ in bulk samples. A key advantage of deriving the Lindblad master equation is that it directly relates the microscopic model to the rates $\gamma_{ij}$, enabling predictions of the expected qubit coherence from the spin-strain interaction strengths.

\subsection{Coherence time in Markovian approximation}

\begin{figure}[ht!]
        \centering
        \includegraphics[width=1\columnwidth]{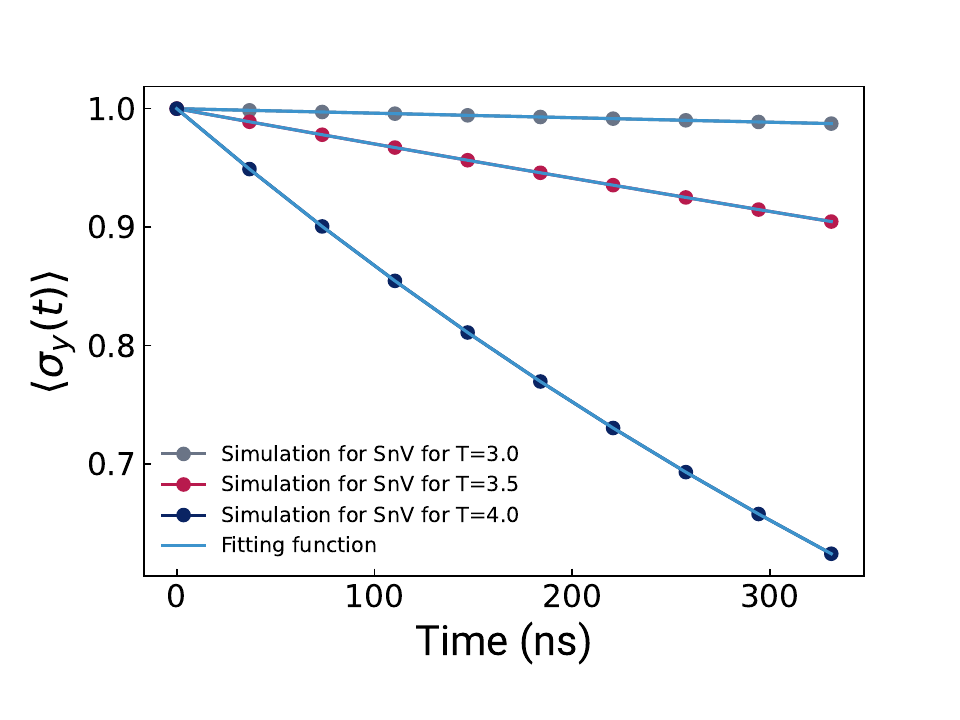}
        \caption{Expectation value $\langle \sigma_y(t)\rangle$ for SnV with orbital splitting $\Delta \approx 903$ GHz with fitting function Eq.~\eqref{eq:Fitting}}
        \label{fig:Fitting function}
\end{figure}

We fit the expectation value $\langle \sigma_y(t)\rangle$ with the function

\begin{equation}
    f(t) = e^{-\frac{t}{T_2}}
\label{eq:Fitting}
\end{equation}
where $T_2$ is the spin coherence time, plotted in Fig.~\ref{fig:Fitting function}. 
We use a fitting function in combination with calculating the time evolution, because at finite temperatures the coherence time cannot be directly determined using just the rates $\gamma_{ij}$ for this multi-level problem. 
In Fig.~\ref{fig:SnV-Th-Exp T_2} we compare both theoretical and experimental values of $T_2$ for SnV as a function of Temperature. 

At low temperature, the measured coherence time is approximately temperature independent because spin-spin interactions and other temperature-independent effects are dominant in this regime. Since our simulations include only the phonon bath, $T_2$ continues to increase at low temperature. At higher temperature, however, the theoretically predicted coherence time becomes much shorter than the experimentally measured value. This is unexpected because the simulations assume ideal conditions and include only spin-phonon coupling, whereas experiments are also affected by spin-spin interactions and other imperfections that should further reduce $T_2$.
\begin{figure}[ht!]
        \centering
        \includegraphics[width=1\columnwidth]{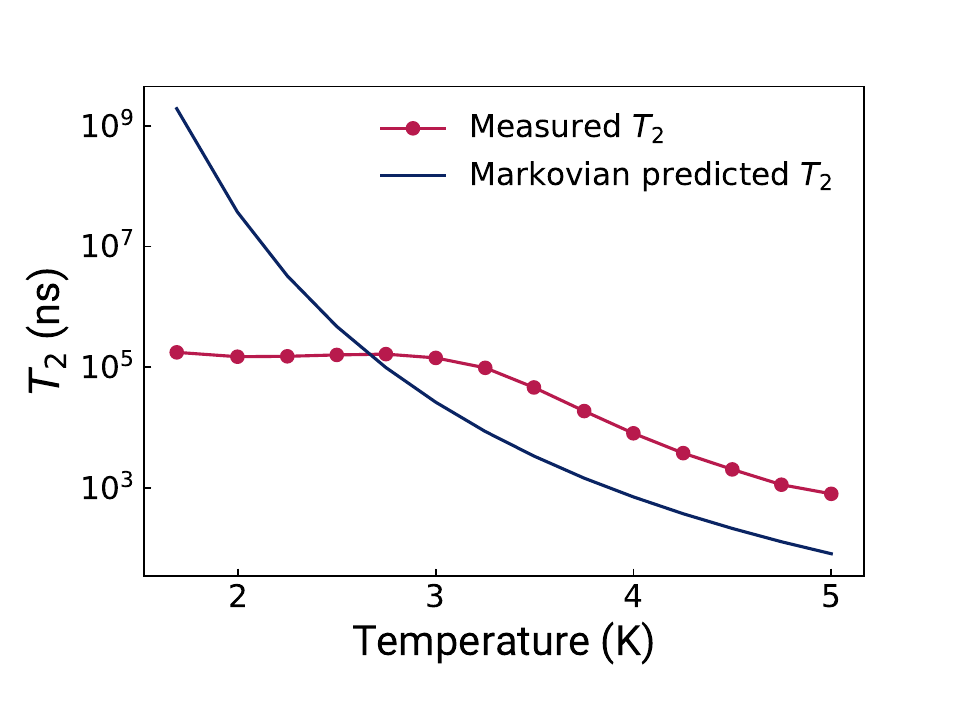}
        \caption{Coherence time $T_2$ for SnV with Hahn-echo as a function of Temperature compared to experiments. Data reported in \cite{harris2024} based on \cite{rosenthal2023}}
        \label{fig:SnV-Th-Exp T_2}
    \end{figure}

This result reveals a limitation of the Markovian model. One shortcoming is that the spectral density is derived from a Debye model for a large bulk sample under ideal conditions. In realistic nanostructures, however, the geometry is more complex and the spectral density may differ \cite{behuninDimensionalTransformationDefectinduced2016}. Moreover, when one or two spatial dimensions are much larger than the others \cite{lemonde2018}, the effective phonon density of states can become two- or one-dimensional.
This can make a signigicant difference for the expected $\gamma_{ij}$ as was shown in \cite{behuninDimensionalTransformationDefectinduced2016}. 

Another important source of the discrepancy between experiment and theory is that inserting the spectral density $J(\omega)=\omega^3$ into Eq.~\eqref{eq:Correlation} yields a divergent correlation function, thereby violating the Markovian approximation. This issue can be overlooked by the use of the identity in Eq.~\eqref{eq:identity}, which removes the integration over the infinite frequency range in the decay rate. Regularization of the spectral density is therefore usually introduced only in the non-Markovian derivation, where integration over the full frequency range is unavoidable.

In previous work \cite{pieplow2024}, the orientation of the static magnetic field was shown to have a crucial effect on microwave control of G4V. Therefore, in our analysis, we consider both orientations: parallel and perpendicular to the symmetry axis of G4V. Experimental measurements of the coherence time are available only for the parallel configuration. However, comparing the two configurations within the Markovian model, both with and without regularization, we find almost no difference in most cases, with the largest deviation remaining below $1\%$.

\subsection{Regularization}

To make sure that the spectral density does not cause a violation of the Markovian approximation, we need to regularize it in the Markovian master equation Eq.~\eqref{eq:NonMarkov}-\eqref{eq:Correlation}.
\begin{figure}[h!]
        \centering
        \includegraphics[width=1\columnwidth]{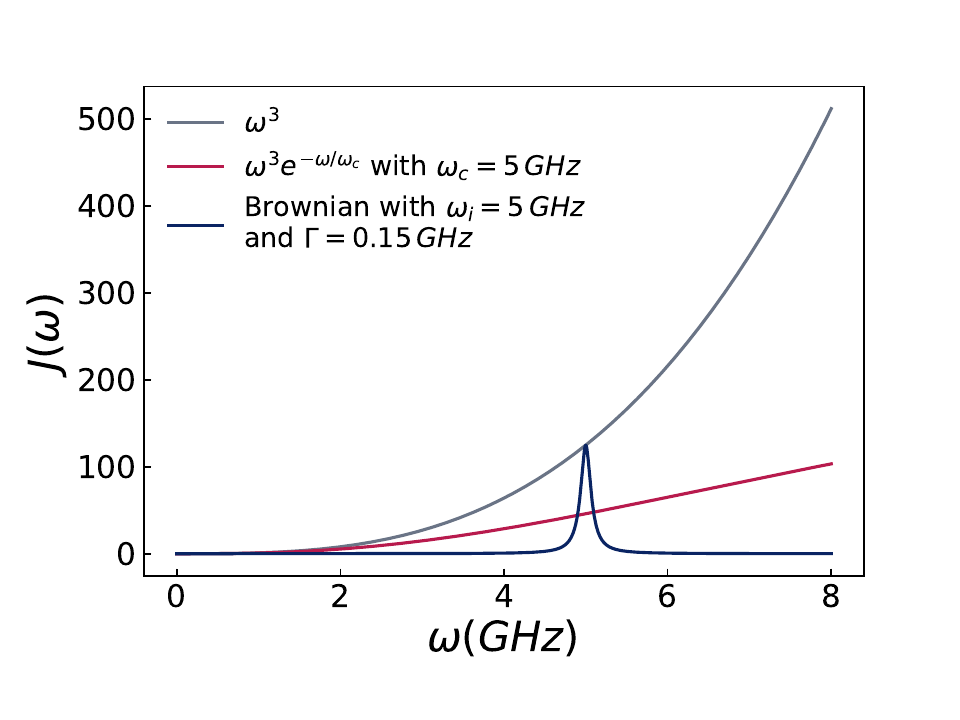}
        \caption{Regularized spectral density. Exponential regularization with bath cutoff frequency $\omega_c = 5$ GHz and Brownian spectral density with distribution linewidth $\Gamma = 0.15$ GHz}
        \label{fig:Reg}
    \end{figure}

{\bf Exponential regularization} We use the Ohmic type spectral density \cite{leggett1987}
\begin{equation}
    J(w) = \chi w^d e^{-w/w_c}
\end{equation}
where $\chi$ is the coupling constant and $w_c$ is the phonon cutoff frequency. In this form, the added exponential decay, will induce convergence in the bath correlation function and result in a reduction of the spin-phonon coupling. 
\begin{figure}[h!]
        \centering
        \includegraphics[width=1\columnwidth]{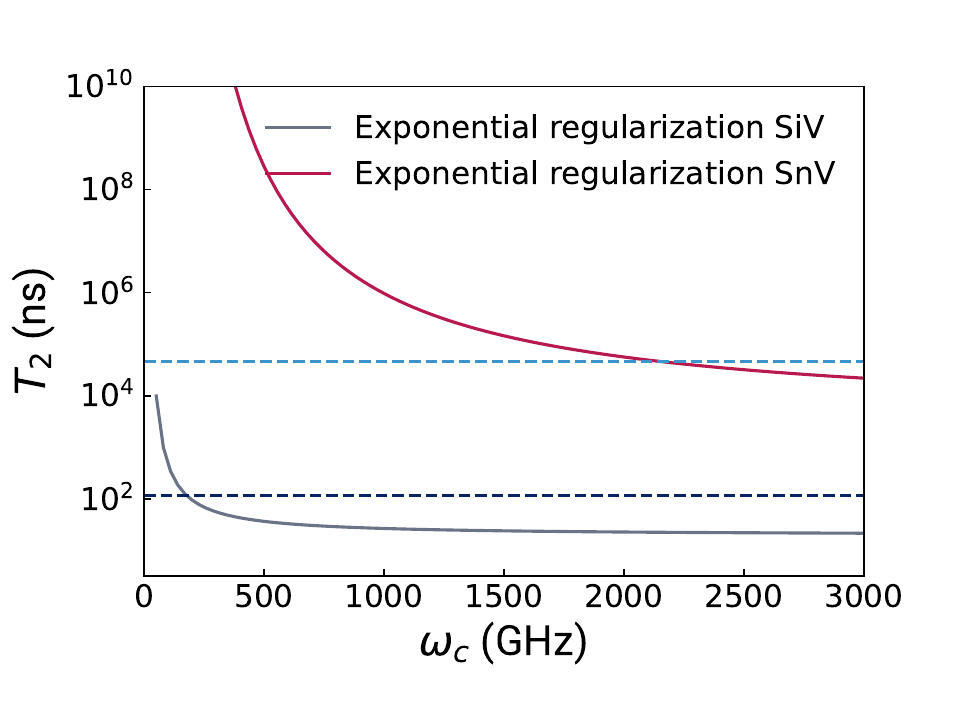}
        \caption{Coherence time $T_2$ of SiV and SnV as a function of $w_c$ compared to experiments. For SiV, measured at $T = 3.6$ K and orbital splitting $\approx 50$ GHz \cite{pingault2017b}. For SnV, measured at $T \approx 3.5$ K and orbital splitting $\Delta\approx 903$ GHz \cite{rosenthal2023}. The dash lines are the experiment measurement for SiV $T_2 = 114.19$ ns and SnV $T_2 = 46$ $\mu$s}
        \label{fig:Mark Exp Reg}
    \end{figure}
    
Interestingly, the calculated coherence time is now mainly determined by the choice of the cutoff frequency $w_c$ in Fig.~\ref{fig:Mark Exp Reg}. Intuitively, two choices can be considered, the Debye frequency $w_D \approx 40$ THz related to the diamond structure as it is the highest frequency allowed in the diamond lattice or the color center orbital splitting, which we motivate in the following: Fig. \ref{fig:Mark Exp Reg} shows that the coherence times for both $SiV$ and $SnV$ are above the measured value around the respective orbital splitting and below it for higher cutoff frequency. This is expected since transitions between two levels are theoretically mediated mostly by field modes close to the orbital resonance. 

Additionally, it was shown experimentally in \cite{kuruma2025} that creating a phononic band gap around the color-center orbital splitting energy can improve the color-center lifetime by more than one order of magnitude. This demonstrates that resonant phonons are the main contributors. Also note that the model under investigation is inherently limited to low frequencies. As discussed in the strain based modeling of the spin-phonon interaction, we are limited to phonons with wavelengths larger than the G4V size, approximately $0.5$ nm for SnV \cite{thieringInitioMagnetoOpticalSpectrum2018}, where the strain based interaction with phonons is expected to be adequate. Near the Debye-frequency range we do not expect meaningful results.
{\bf Brownian spectral density} originates from modeling the environment as underdamped harmonic oscillator linearly coupled to the system \cite{lambert2023}. It is a more physically motivated choice than exponential regularization, as it naturally preserves the coupling strength while also resolving limitations of the Born approximation as it implies thar the phonons evolve and decay. In Eq.~\eqref{eq: Lindblad}, only the resonant frequencies of the spectral density contribute. We therefore choose a Brownian spectral density to extend the model to a broader distribution around the resonance frequency.
\begin{equation}
    J(\omega) = \sum_i \frac{\alpha^2 \Gamma \, \omega}{\left[(\omega_i^2 - \omega^2)^2 + \Gamma^2 \omega^2\right]}
\end{equation}
where $w_i$ are the color centers transition frequencies and $\Gamma$ is the linewidth of the distribution. To keep the spin-phonon coupling unchanged in the Markovian master equation, we set $\alpha^2 = \Gamma w_i^2$. This spectral density reproduces Eq.~\eqref{eq: Lindblad}. Moreover, inserting it into Eq.~\eqref{eq:Correlation} shows that $\Gamma$ is also the decay rate of the bath correlation function, so that the Markovian approximation is justified.
\begin{equation}
\label{eq:BrownianCorrelation}
    \left\langle B_\alpha(t)^\dagger\,B_\beta(t - s) \right\rangle \propto e^{-\Gamma s/2}
\end{equation}
This spectral density is well suited to our analysis: it leaves the Markovian results unchanged while introducing the bath memory required for non-Markovian effects. It also avoids introducing cutoff frequencies to regularize the spectral density, at the cost of an effective bath memory time that cannot be predicted from diamond phonons without accounting for e.g. interphonon interactions.

\section{Non-Markovianity}

It is clear from Eq.~\eqref{eq: Lindblad} that within the Markovian approximation the diagonal part of the spin Hamiltonian does not contribute to the system evolution. However, interactions that are diagonal in the system,
\begin{equation}
    H_{SB} = \sum_{k} \sigma_{z} \left( g_k b_k^{\dagger} + g_k^{*} b_k \right),
\end{equation}
are known to be exactly solvable \cite{breuerTheoryOpenQuantum2009} and can modify the observed coherences. These shortcomings motivate the extension of our simulations to the non-Markovian regime. 

For the non-Markovian simulations, we consider the integro-differential equation, Eq.~\eqref{eq:Master equation}, under the Born approximation, both with and without the RWA. The latter case is numerically more demanding and can only be simulated over short time intervals. In addition, we consider the simplified TCL equation, Eq.~\eqref{eq:NonMarkov}, which includes only bath-memory effects and, in general, does not preserve positivity unless the RWA is applied. We again consider parallel and perpendicular magnetic-field configurations. For both orientations, the exponential regularization yields a maximum difference of only $1.2\%$. For the Brownian spectral density, however, the parallel configuration gives a $26\%$ larger $T_2$ in the TCL approach and a $19$--$46\%$ larger $T_2$ in the IDE approach. This observation may enable a test of the theoretical models by measuring the spin-qubit coherence time for two different magnetic-field orientations, providing a relatively simple way to exclude a theoretical model.

\subsection{Discussion of results}
For the non-Markovian analysis, we now use a more general exponential decay function,
\begin{equation}
    f_p(t) = e^{-\left(\frac{t}{T_2}\right)^p},
\end{equation}
where $p\in [1,2]$. Also at very short time, the behavior can be different from the exponential decay. Unfortunately,  simulations cannot be performed for all points and all cases, as they may fail due to the inherent instability of Eq.~\eqref{eq:Master equation}, numerical divergence at high frequencies, or limitations in computational time.

Furthermore, due to the exponential terms in the non-Markovian equations, the SnV simulations become unstable. This behavior is partly due to the intrinsic instability of the equations and partly to high oscillating terms.

{\bf Exponential regularization} We can obtain an expression for the decay rate in Eq.~\eqref{eq:NonMarkov} by using an expansion of the Bose-Einstein distribution:
\begin{equation}
    n(w) = \sum_{{n=1}}^{{{N}}} e^{{-n \frac{\hbar w}{k_B T}}}
\end{equation}
This also reduces numerical instabilities that occur when performing the double integration over frequency and time, while decreasing the computational cost. The non-Markovian contribution is almost negligible, with coherence times differing by less than $1.3\%$ from the Markovian simulations.

\begin{figure}[h!]
        \centering
        \includegraphics[width=1\columnwidth]{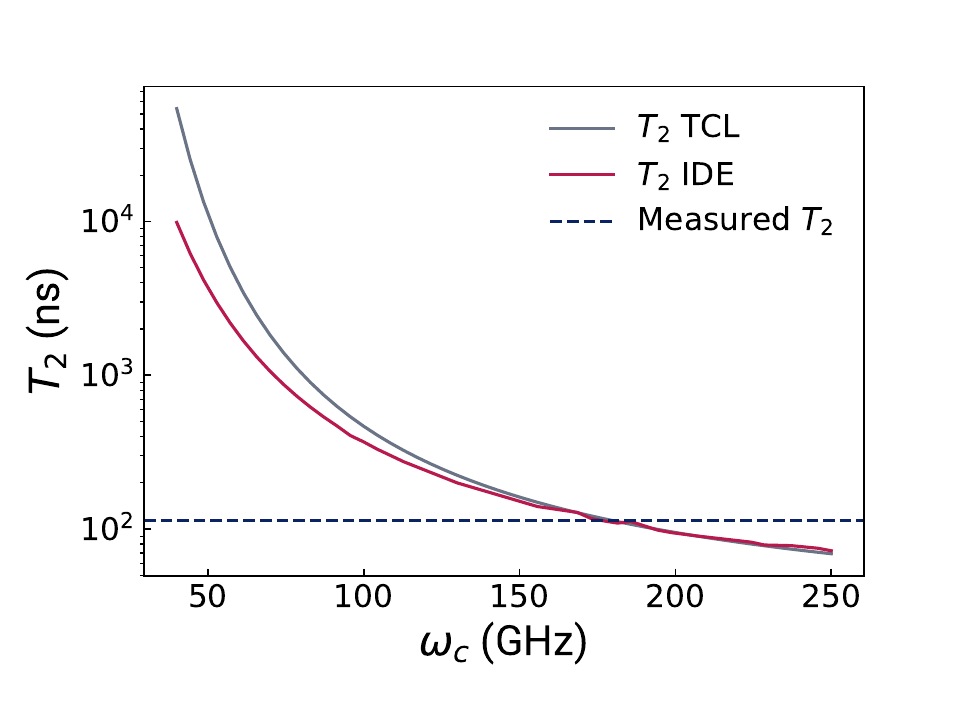}
        \caption{Coherence time $T_2$ of SiV at $T = 3.6$ K and orbital splitting $\Delta\approx 50$ GHz \cite{pingault2017b} as a function of $w_c$ using TCL Eq.~\eqref{eq:NonMarkov} and the IDE Eq.~\eqref{eq:Master equation}. The dash line is the experiment measurement for SiV $T_2 = 114.19$ ns}
        \label{fig:NM Exp}
\end{figure}

\begin{figure}[h!]
        \centering
        \includegraphics[width=1\columnwidth]{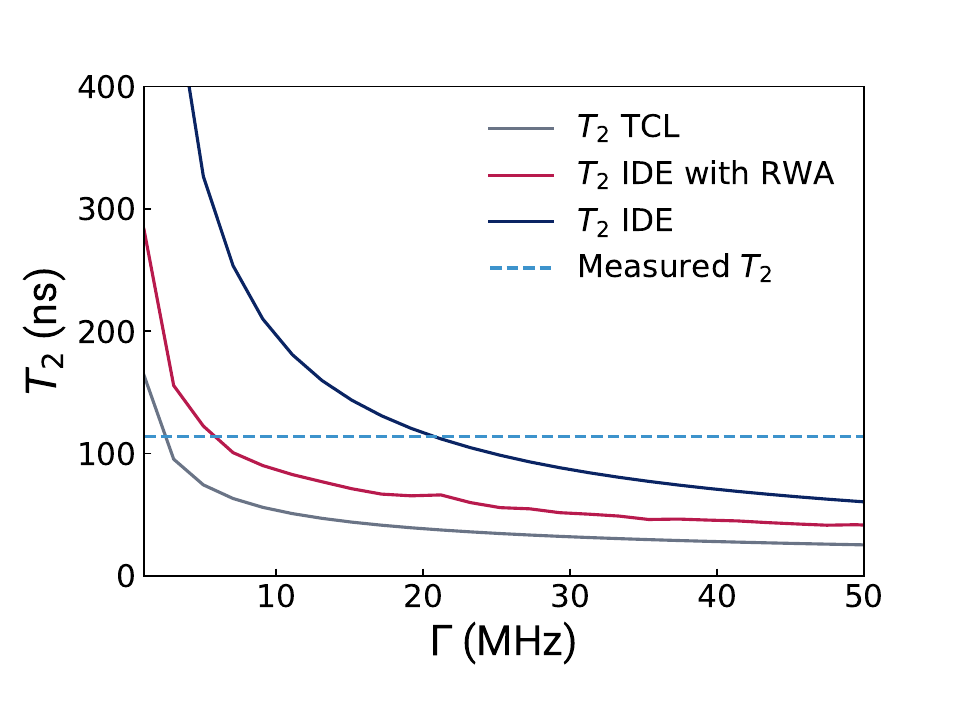}
        \caption{Coherence time $T_2$ of SiV at $T = 3.6$ K and orbital splitting $\Delta\approx 50$ GHz \cite{pingault2017b} as a function of $\Gamma$ using TCL Eq.~\eqref{eq:NonMarkov} and the IDE Eq.~\eqref{eq:Master equation}. The dashed line shows the measured coherence time for the SiV $T_2 = 114.19$ ns}
        \label{fig:NM Brow}
    \end{figure}

The integration of Eq.~\eqref{eq:Master equation} involves only a frequency integral. In this particular case, an exact solution exists for the bath correlation function \cite{lambert2023, Brandes}
\begin{widetext}
\begin{equation}
\label{eq:CorrelationExp}
    \left\langle B_\alpha(t)^\dagger\,B_\beta(t - s) \right\rangle = \delta_{\alpha\beta}\chi\,\beta^{-(s+1)} \Gamma(s+1)
\left[
\zeta\!\left(s+1, \frac{1+\beta \omega_c -  {\rm i} \omega_c t}{\beta \omega_c}\right)
+
\zeta\!\left(s+1, \frac{1+ {\rm i} \omega_c t}{\beta \omega_c}\right)
\right]
\end{equation}
\end{widetext}
where $\Gamma$ is the Gamma function, $\zeta$ is the generalized zeta function and $\beta = \hbar/K_B T$. The non-Markovianity here is very distinctive, where including the system memory remarkably reduces the calculated coherence time as plotted in Fig.~\ref{fig:NM Exp}. The reduction can be explained by the fact that in the TCL approach, an approximation of the correlation function was used. Additionally, its short-time behavior, marked by fast relaxation in the time evolution, differs remarkably from its long term behavior, which becomes apparent once the system memory is included.
{\bf Brownian spectral density} This spectral density is particularly well suited for analyzing non-Markovianity, as it provides a regularization that holds for the Markovian approximation while simultaneously introducing a finite bath memory. Additionally, the bath correlation function is exactly solvable and is written in terms of decaying exponential functions \cite{lambert2023}
\begin{equation}
\label{eq:CorrelationBrow}
\begin{aligned}
    \left\langle B_\alpha(t)^\dagger\,B_\beta(t - s) \right\rangle &= \delta_{\alpha\beta} \times \\ 
    & \hspace{-1.5cm}\chi \sum_{k=0}^\infty (C_k^R e^{-i\nu_k^R s} + i C_k^I e^{-i\nu_k^I s})
\end{aligned}
\end{equation}
\small
\begin{align}
C_k^{R} &=
\begin{cases}
\dfrac{\alpha^{2} \coth\!\left(\beta(\Omega +  {\rm i}\Gamma/2)/2\right)}{4\Omega}, & k = 0 \\[8pt]
\dfrac{\alpha^{2} \coth\!\left(\beta(\Omega -  {\rm i}\Gamma/2)/2\right)}{4\Omega}, & k = 0 \\[10pt]
\dfrac{-2\alpha^{2}\Gamma \epsilon_k}
{\beta \left[(\Omega +  {\rm i}\Gamma/2)^2 + \epsilon_k^2\right]
\left[(\Omega -  {\rm i}\Gamma/2)^2 + \epsilon_k^2\right]}, 
& k \ge 1
\end{cases}\\
\nu_k^{R} &=
\begin{cases}
- {\rm i}\Omega + \Gamma/2, & k = 0 \\
 {\rm i}\Omega + \Gamma/2, & k = 0 \\
2\pi k/\beta, & k \ge 1
\end{cases}\\
C_k^{I} &=
\begin{cases}
 {\rm i}\alpha^{2}/4\Omega, & k = 0 \\
- {\rm i}\alpha^{2}/4\Omega, & k = 0
\end{cases}\\
\nu_k^{I} &=
\begin{cases}
 {\rm i}\Omega + \Gamma/2, & k = 0 \\
- {\rm i}\Omega + \Gamma/2, & k = 0
\end{cases}
\end{align}
\normalsize
where $\Omega = \sqrt{\omega_i^{2} - (\Gamma/2)^2}$ and $\epsilon_k = 2\pi k/\beta$. $\Gamma$ plays the role of the decay rate of the bath memory Eq.~\eqref{eq:BrownianCorrelation}. Fig.~\ref{fig:NM Brow} shows that the coherence time $T_2$ increases for lower bath decay rates which is expected since the information flows back and forth between the system and the bath before it is lost due to the bath decay. Also, when we consider the system memory via the IDE Eq.~\eqref{eq:Master equation}, coherence times further increase from $2$ to $6.7$ times. Under the RWA the increase is limited to around $1.7$ times, since the approximation discards part of the information flow.

\section{Conclusion}

In this work, we highlight the difficulty of predicting G4V electronic spin coherence times accurately and consistently from first principles. Such predictions are essential for assessing intrinsic platform limitations under idealized conditions. Even beyond this limit, microscopic modeling can guide bath-engineering strategies, for example by reducing the effective bath dimensionality \cite{lemonde2018}or by introducing gaps in the bath spectral density \cite{kuruma2025}. 

In the case of G4Vs, the challenge begins with the choice of a spin-phonon model, but persists even in the frequency regime where a strain-based description is appropriate. In this regime, ambiguities remain in the treatment of bath memory and, more importantly, in the spectral properties of the phonon bath.

The Markovian master equation provides a controlled description in the weak-coupling regime, but it has important limitations. These limitations directly affect the predicted coherence times, especially when the theory unexpectedly yields a shorter $T_2$ than measured experimentally. We trace this discrepancy to several assumptions: the standard Markovian treatment contains divergent terms that require renormalization, removes pure-dephasing contributions from the interaction Hamiltonian, and often relies on a Debye spectral density that may be too restrictive for realistic phonon baths. We showed that the spectral density can substantially influence the predicted coherence time and that additional bath parameters may be required for quantitative agreement.

Non-Markovian corrections are likewise sensitive to the spectral density of the phonon bath. They are more apparent for the Brownian spectral density than for exponential regularization. We compare two non-Markovian descriptions: one including only bath memory and another additionally incorporating system memory. In both cases, memory effects prolong the calculated coherence time by allowing information to flow back and forth between the system and the bath before it is irreversibly lost.

Finally, we identified a simple experimental test that could discriminate between the two spectral density models. Measuring the electronic spin coherence time of a SiV or SnV center for two different static magnetic-field orientations should, in principle, be sufficient to rule out the exponential-regularization method combined with a Debye spectral density. This test requires that other decoherence channels be strongly suppressed, for example spin noise, which can be largely removed in isotopically purified diamond \cite{balasubramanian2009a}, and surface-induced noise, which can be reduced by using isolated vacancies deeply implanted into the diamond \cite{romach2015}.

\begin{acknowledgments}

Funding for this project was provided by the German Federal Ministry of Education and Research (BMBFTR, project QPIS, No. 16KISQ032K; project QPIS.2, No. KIS6GCQ020; project DINOQUANT 13N14921, ERC StG project QUREP of the EC, No. 851810).

\end{acknowledgments}

\section*{Author Contributions}

M.B. and G.P. conceptualized the research and developed the core idea. M.B conducted the calculations and implemented the numerical simulations and data analysis. T.S. provided overall project supervision. All authors contributed to writing and refining the manuscript. ChatGPT has been used for grammar checking and bibliographic research.
\bibliographystyle{quantum}
\bibliography{Bib.bib}

@article{singh_modular_2025,
	title = {\textit{Modular architectures and entanglement schemes for error-corrected distributed quantum computation}},
	volume = {12},
	url = {https://www.nature.com/articles/s41534-025-01146-2},
	doi = {10.1038/s41534-025-01146-2},
	pages = {},
	number = {1},
	journal = {npj Quantum Information},
	author = {Singh, Siddhant and Gu, Fenglei and de Bone, Sébastian and Villaseñor, Eduardo and Elkouss, David and Borregaard, Johannes},
	year = {2025},
}

@article{abobeih2022a,
	title = {\textit{Fault-Tolerant Operation of a Logical Qubit in a Diamond Quantum Processor}},
	volume = {606},
	url = {https://www.nature.com/articles/s41586-022-04819-6},
	doi = {10.1038/s41586-022-04819-6},
	pages = {884-889},
	number = {7916},
	journal = {Nature},
	author = {Abobeih, M. H. and Wang, Y. and Randall, J. and Loenen, S. J. H. and Bradley, C. E. and Markham, M. and Twitchen, D. J. and Terhal, B. M. and Taminiau, T. H.},
	year = {2022},
}

@article{ruf2021,
	title = {\textit{Quantum Networks Based on Color Centers in Diamond}},
	volume = {130},
	url = {https://doi.org/10.1063/5.0056534},
	doi = {10.1063/5.0056534},
	pages = {070901},
	number = {7},
	journal = {Journal of Applied Physics},
	author = {Ruf, Maximilian and Wan, Noel H. and Choi, Hyeongrak and Englund, Dirk and Hanson, Ronald},
	year = {2021},
}

@article{pezzagna2021,
	title = {\textit{Quantum Computer Based on Color Centers in Diamond}},
	volume = {8},
	url = {https://doi.org/10.1063/5.0007444},
	doi = {10.1063/5.0007444},
	pages = {011308},
	number = {1},
	journal = {Applied Physics Reviews},
	author = {Pezzagna, S{\'e}bastien and Meijer, Jan},
	year = {2021},
}

@article{zotero-item-5989,
	title = {\textit{Cryogenic platform for coupling color centers in diamond membranes to a fiber-based microcavity}},
	volume = {126},
	url = {https://link.springer.com/article/10.1007/s00340-020-07478-5},
	doi = {https://doi.org/10.1007/s00340-020-07478-5},
	pages = {},
	number = {},
	journal = {Applied Physics B},
	author = {M. Salz and Y. Herrmann and A. Nadarajah, A. Stahl and M. Hettrich and A. Stacey and S. Prawer and D. Hunger and F. Schmidt-Kaler},
	year = {2020},
}

@article{becker2018,
	title = {\textit{All-{{Optical Control}} of the {{Silicon-Vacancy Spin}} in {{Diamond}} at {{Millikelvin Temperatures}}}},
	volume = {120},
	url = {https://doi.org/10.1103/PhysRevLett.120.053603},
	doi = {10.1103/PhysRevLett.120.053603},
	pages = {},
	number = {},
	journal = {Physical Review Letters},
	author = {Becker, Jonas N. and Pingault, Benjamin and Gro{\ss}, David and G{\"u}ndo{\u g}an, Mustafa and Kukharchyk, Nadezhda and Markham, Matthew and Edmonds, Andrew and Atat{\"u}re, Mete and Bushev, Pavel and Becher, Christoph},
	year = {2018},
}

@article{knaut2024entanglement,
  author  = {Knaut, C. M. and Suleymanzade, A. and Wei, Y.-C. and Assumpcao, D. R. and Stas, P.-J. and Huan, Y. Q. and Machielse, B. and Knall, E. N. and Sutula, M. and Baranes, G. and Sinclair, N. and De-Eknamkul, C. and Levonian, D. S. and Bhaskar, M. K. and Park, H. and Lon{\v{c}}ar, M. and Lukin, M. D.},
  title   = {\textit{Entanglement of nanophotonic quantum memory nodes in a telecom network}},
  journal = {Nature},
  volume  = {629},
  number  = {8012},
  pages   = {573--578},
  year    = {2024},
  doi     = {10.1038/s41586-024-07252-z},
  url     = {https://www.nature.com/articles/s41586-024-07252-z}
}

@article{wei2025,
	title = {\textit{Universal Distributed Blind Quantum Computing with Solid-State Qubits}},
	volume = {388},
	url = {https://www.science.org/doi/10.1126/science.adu6894},
	doi = {10.1126/science.adu6894},
	pages = {},
	number = {509-513},
	journal = {Science},
	author = {Wei, Y.-C. and Stas, P.-J. and Suleymanzade, A. and Baranes, G. and Machado, F. and Huan, Y. Q. and Knaut, C. M. and Ding, S. W. and Merz, M. and Knall, E. N. and Yazlar, U. and Sirotin, M. and Wang, I. W. and Machielse, B. and Yelin, S. F. and Borregaard, J. and Park, H. and Lon{\v c}ar, M. and Lukin, M. D.},
	year = {2025},
}

@article{heppElectronicStructureSilicon2014,
	title={\textit{Electronic structure of the silicon vacancy color center in diamond}}, 
        author={Hepp, Christian}, 
        journal = {doi:10.22028/D291-23020},
        url={http://dx.doi.org/10.22028/D291-23020}, 
        year={2014},
}

@article{londero2018,
	title = {\textit{Vibrational Modes of Negatively Charged Silicon-Vacancy Centers in Diamond from Ab Initio Calculations}},
	volume = {98},
	url = {https://doi.org/10.1103/PhysRevB.98.035306},
	doi = {10.1103/PhysRevB.98.035306},
	pages = {},
	number = {},
	journal = {Physical Review B},
	author = {Londero, Elisa and Thiering, Gergő and Razinkovas, Lukas and Gali, Adam and Alkauskas, Audrius},
	year = {2018},
}

@article{harris2023,
	title = {\textit{Coherence of Group-{{IV}} Color Centers}},
	volume = {109},
	url = {https://doi.org/10.1103/PhysRevB.109.085414},
	doi = {10.1103/PhysRevB.109.085414},
	pages = {},
	number = {},
	journal = {Physical Review B},
	author = {Harris, Isaac B. W. and Englund, Dirk},
	year = {2024},
}

@book{breuerTheoryOpenQuantum2009,
	title = {\textit{The Theory of Open Quantum Systems}},
	volume = {},
	url = {/10.1093/acprof:oso/9780199213900.001.0001},
	doi = {10.1093/acprof:oso/9780199213900.001.0001},
	pages = {},
	number = {},
	publisher = {Oxford University Press},
	author = {Breuer, Heinz-Peter and Petruccione, Francesco},
	year = {2009},
}

@article{behuninDimensionalTransformationDefectinduced2016,
	title = {\textit{Dimensional Transformation of Defect-Induced Noise, Dissipation, and Nonlinearity}},
	volume = {93},
	url = {https://doi.org/10.1103/PhysRevB.93.224110},
	doi = {10.1103/PhysRevB.93.224110},
	pages = {},
	number = {},
	journal = {Physical Review B},
	author = {Behunin, R. O. and Intravaia, F. and Rakich, P. T.},
	year = {2016},
}

@article{lemonde2018,
	title = {\textit{Phonon {{Networks}} with {{Silicon-Vacancy Centers}} in {{Diamond Waveguides}}}},
	volume = {120},
	url = {https://link.aps.org/doi/10.1103/PhysRevLett.120.213603},
	doi = {10.1103/PhysRevLett.120.213603},
	pages = {213603},
	number = {21},
	journal = {Phys. Rev. Lett},
	author = {Lemonde, M.-A. and Meesala, S. and Sipahigil, A. and Schuetz, M. J. A. and Lukin, M. D. and Loncar, M. and Rabl, P.},
	year = {2018},
}

@article{leggett1987,
	title = {\textit{Dynamics of the Dissipative Two-State System}},
	volume = {59},
	url = {https://link.aps.org/doi/10.1103/RevModPhys.59.1},
	doi = {10.1103/RevModPhys.59.1},
	pages = {},
	number = {1},
	journal = {Rev. Mod. Phys.},
	author = {Leggett, A. J. and Chakravarty, S. and Dorsey, A. T. and Fisher, Matthew P. A. and Garg, Anupam and Zwerger, W.},
	year = {1987},
}

@article{kuruma2025,
	title = {\textit{Controlling Interactions between High-Frequency Phonons and Single Quantum Systems Using Phononic Crystals}},
	volume = {21},
	url = {https://www.nature.com/articles/s41567-024-02697-5},
	doi = {10.1038/s41567-024-02697-5},
	pages = {77-82},
	number = {1},
	journal = {Nat. Phys.},
	author = {Kuruma, Kazuhiro and Pingault, Benjamin and Chia, Cleaven and Haas, Michael and Joe, Graham D. and Assumpcao, Daniel Rimoli and Ding, Sophie Weiyi and Jin, Chang and Xin, C. J. and Yeh, Matthew and Sinclair, Neil and Lončar, Marko},
	year = {2025},
}

@article{schaller2008,
	title = {\textit{Preservation of Positivity by Dynamical Coarse Graining}},
	volume = {78},
	url = {https://link.aps.org/doi/10.1103/PhysRevA.78.022106},
	doi = {10.1103/PhysRevA.78.022106},
	pages = {022106},
	number = {2},
	journal = {Phys. Rev. A},
	author = {Schaller, Gernot and Brandes, Tobias},
	year = {2008},
}

@article{pieplow2024,
	title = {\textit{Efficient {{Microwave Spin Control}} of {{Negatively Charged Group-IV Color Centers}} in {{Diamond}}}},
	volume = {109},
	url = {https://doi.org/10.1103/PhysRevB.109.115409},
	doi = {10.1103/PhysRevB.109.115409},
	pages = {115409},
	number = {11},
	journal = {Physical Review B},
	author = {Pieplow, Gregor and Belhassen, Mohamed and Schr{\"o}der, Tim},
	year = {2024},
}

@software{harris2024,
	title = {\textit{Ibwharri/{{Coherence-of-Group-IV-Color-Centers}}}},
	volume = {},
	url = {https://github.com/ibwharri/Coherence-of-Group-IV-Color-Centers},
	doi = {https://github.com/ibwharri/Coherence-of-Group-IV-Color-Centers},
	pages = {},
	number = {},
	journal = {Github},
	author = {Harris, Isaac B. W. and Englund, Dirk},
	year = {2023},
}

@article{rosenthal2023,
	title = {\textit{Microwave {{Spin Control}} of a {{Tin-Vacancy Qubit}} in {{Diamond}}}},
	volume = {13},
	url = {https://link.aps.org/doi/10.1103/PhysRevX.13.031022},
	doi = {10.1103/PhysRevX.13.031022},
	pages = {031022},
	number = {3},
	journal = {Phys. Rev. X},
	author = {Rosenthal, Eric I. and Anderson, Christopher P. and Kleidermacher, Hannah C. and Stein, Abigail J. and Lee, Hope and Grzesik, Jakob and Scuri, Giovanni and Rugar, Alison E. and Riedel, Daniel and Aghaeimeibodi, Shahriar and Ahn, Geun Ho and Van Gasse, Kasper and Vučković, Jelena},
	year = {2023},
}

@article{pingault2017b,
	title = {\textit{Coherent Control of the Silicon-Vacancy Spin in Diamond}},
	volume = {8},
	url = {https://www.nature.com/articles/ncomms15579},
	doi = {10.1038/ncomms15579},
	pages = {15579},
	number = {1},
	journal = {Nature Communications},
	author = {Pingault, Benjamin and Jarausch, David-Dominik and Hepp, Christian and Klintberg, Lina and Becker, Jonas N. and Markham, Matthew and Becher, Christoph and Atat{\"u}re, Mete},
	year = {2017},
}

@article{lambert2023,
	title = {\textit{{{QuTiP-BoFiN}}: {{A}} Bosonic and Fermionic Numerical Hierarchical-Equations-of-Motion Library with Applications in Light-Harvesting, Quantum Control, and Single-Molecule Electronics}},
	volume = {5},
	url = {https://doi.org/10.1103/PhysRevResearch.5.013181},
	doi = {10.1103/PhysRevResearch.5.013181},
	pages = {013181},
	number = {1},
	journal = {Physical Review Research},
	author = {Lambert, Neill and Raheja, Tarun and Cross, Simon and Menczel, Paul and Ahmed, Shahnawaz and Pitchford, Alexander and Burgarth, Daniel and Nori, Franco},
	year = {2023},
}

@article{Brandes,
	title = {\textit{Lectures on Background to Quantum Information}},
	volume = {},
	url = {https://www1.itp.tu-berlin.de/brandes/public_html/publications/notes.pdf},
	doi = {},
	pages = {},
	number = {},
	journal = {Chapter 7 (Quantum dissipation)},
	author = {T. Brandes},
	year = {2003},
}

@article{meesala2018,
	title = {\textit{Strain Engineering of the Silicon-Vacancy Center in Diamond}},
	volume = {97},
	url = {https://link.aps.org/doi/10.1103/PhysRevB.97.205444},
	doi = {10.1103/PhysRevB.97.20544},
	pages = {205444},
	number = {20},
	journal = {Phys. Rev. B},
	author = {Meesala, Srujan and Sohn, Young-Ik and Pingault, Benjamin and Shao, Linbo and Atikian, Haig A. and Holzgrafe, Jeffrey and Gündoğan, Mustafa and Stavrakas, Camille and Sipahigil, Alp and Chia, Cleaven and Evans, Ruffin and Burek, Michael J. and Zhang, Mian and Wu, Lue and Pacheco, Jose L. and Abraham, John and Bielejec, Edward and Lukin, Mikhail D. and Atatüre, Mete and Lončar, Marko},
	year = {2018},
}

@article{bopp2024,
	title = {\textit{'{{Sawfish}}' {{Photonic Crystal Cavity}} for {{Near-Unity Emitter-to-Fiber Interfacing}} in {{Quantum Network Applications}}}},
	volume = {12},
	url = {https://doi.org/10.1002/adom.202301286},
	doi = {10.1002/adom.202301286},
	pages = {2301286},
	number = {13},
	journal = {Advanced Optical Materials},
	author = {Bopp, Julian M. and Plock, Matthias and Turan, Tim and Pieplow, Gregor and Burger, Sven and Schröder, Tim},
	year = {2024},
}

@article{trusheim_transform-limited_2020,
	title = {\textit{Transform-Limited Photons From a Coherent Tin-Vacancy Spin in Diamond}},
	volume = {124},
	url = {https://link.aps.org/doi/10.1103/PhysRevLett.124.023602},
	doi = {10.1103/PhysRevLett.124.023602},
	pages = {023602},
	number = {2},
	journal = {Phys. Rev. Lett.},
	author = {Trusheim, Matthew E. and Pingault, Benjamin and Wan, Noel H. and Gündoğan, Mustafa and De Santis, Lorenzo and Debroux, Romain and Gangloff, Dorian and Purser, Carola and Chen, Kevin C. and Walsh, Michael and Rose, Joshua J. and Becker, Jonas N. and Lienhard, Benjamin and Bersin, Eric and Paradeisanos, Ioannis and Wang, Gang and Lyzwa, Dominika and Montblanch, Alejandro R-P. and Malladi, Girish and Bakhru, Hassaram and Ferrari, Andrea C. and Walmsley, Ian A. and Atatüre, Mete and Englund, Dirk},
	year = {2020},
_note= {American Physical Society},
}

@article{balasubramanian2009a,
  title = {\textit{Ultralong Spin Coherence Time in Isotopically Engineered Diamond}},
  author = {Balasubramanian, Gopalakrishnan and Neumann, Philipp and Twitchen, Daniel and Markham, Matthew and Kolesov, Roman and Mizuochi, Norikazu and Isoya, Junichi and Achard, Jocelyn and Beck, Johannes and Tissler, Julia and Jacques, Vincent and Hemmer, Philip R. and Jelezko, Fedor and Wrachtrup, Jörg},
  year = {2009},
  journal = {Nature Materials},
  volume = {8},
  number = {5},
  pages = {383--387},
  doi = {10.1038/nmat2420},
  url = {https://www.nature.com/articles/nmat2420}
}

@article{romach2015,
  title = {\textit{Spectroscopy of {{Surface-Induced Noise Using Shallow Spins}} in {{Diamond}}}},
  author = {Romach, Y. and Müller, C. and Unden, T. and Rogers, L. J. and Isoda, T. and Itoh, K. M. and Markham, M. and Stacey, A. and Meijer, J. and Pezzagna, S. and Naydenov, B. and McGuinness, L. P. and Bar-Gill, N. and Jelezko, F.},
  year = {2015},
  journal = {Physical Review Letters},
  volume = {114},
  number = {1},
  pages = {017601},
  issn = {0031-9007, 1079-7114},
  doi = {10.1103/PhysRevLett.114.017601},
  url = {https://link.aps.org/doi/10.1103/PhysRevLett.114.017601}
}

@article{thieringInitioMagnetoOpticalSpectrum2018,
  title = {Ab Initio Magneto-Optical Spectrum of Group-IV Vacancy Color Centers in Diamond},
  author = {Thiering, Gergő and Gali, Adam},
  year = {2018},
  journal = {Physical Review X},
  volume = {8},
  number = {2},
  pages = {021063},
  issn = {2160-3308},
  doi = {10.1103/PhysRevX.8.021063},
  url = {https://link.aps.org/doi/10.1103/PhysRevX.8.021063}
}

\end{document}